# Dependencies and systemic risk in the European insurance sector: Some new evidence based on copula-DCC-GARCH model and selected clustering methods

**Stanisław Wanat, Anna Denkowska**

Cracow University of Economics
Faculty of Finance and Law
Department of Mathematics

December 30th 2018

*Abstract*

The subject of the present article is the study of correlations between large insurance companies and their contribution to systemic risk in the insurance sector. Our main goal is to analyze the conditional structure of the correlation on the European insurance market and to compare systemic risk in different regimes of this market. These regimes are identified by monitoring the weekly rates of returns of eight of the largest insurers (five from Europe and the biggest insurers from the USA, Canada and China) during the period January 2005 to December 2018. To this aim we use statistical clustering methods for time units (weeks) to which we assigned the conditional variances obtained from the estimated copula-DCC-GARCH model. The advantage of such an approach is that there is no need to assume a priori a number of market regimes, since this number has been identified by means of clustering quality validation. In each of the identified market regimes we determined the commonly now used *CoVaR* systemic risk measure. From the performed analysis we conclude that all the considered insurance companies are positively correlated and this correlation is stronger in times of turbulences on global markets which shows an increased exposure of the European insurance sector to systemic risk during crisis. Moreover, in times of turbulences on global markets the value level of the *CoVaR* systemic risk index is much higher than in `normal conditions'.

## 1. Introduction

Following the financial crisis in the years 2007-2009 and the European public debt crisis in the years 2010-2012, there has been significantly growing interest in systemic risk. This resulted in a prolific specialized literature proposing, among others, a wide range of new methods for the study of the influence of financial institutions on systemic risk. Moreover, both the academic community and the financial regulatory authorities paid more attention to the role played by non-bank financial institutions, in particular insurance companies, in creating systemic risk. Before the crisis it was generally accepted that the insurance market has a negligible impact on it. However, in the subsequent literature – although many a study still supported the latter point of view – there appeared several articles suggesting the possibility of the insurance market creating systemic risk. Let us quote here a few articles the authors of which claim that insurance companies:

− have become an unavoidable source of systemic risk (e.g. [Billio et al.. 2012] , [Weiß, Mühlnickel 2014]);

− can be systemically important, but only due to their non-traditional (banking) activities (e.g. [Baluch et al.. 2011], [Cummins, Weiss 2014]), while in general the systemic significance of the insurance sector as a whole is still subordinated to the banking sector ([Chen et al. 2013], [Czerwińska 2014]);

− are systemically unimportant due to the low level of interconnections and the lack of a strong dependance on external funding (e.g. [Harrington 2009], [Bell, Keller 2009], [Geneva Association 2010], [Bednarczyk 2013]).

On the other hand, in [Bierth et al. 2015] the authors, after having studied a very large sample of insurers in a long-term horizon, claim that the contribution of the insurance sector to systemic risk is relatively small, its peak having been reached during the financial crisis in the years 2007-2008. They also indicate the four L's: linkages, leverage, losses, liquidity, as the crucial factors influencing the exposure of insurers to systemic risk.

    The present article belongs to the main stream of the studies concerning the linkages between large insurance companies and their contribution to systemic risk in the insurance sector. Our main aim is to check whether the strength of the existing connections between the eight largest insurers (five from Europe, one from the USA, Canada and China) together with their contribution to systemic risk in the European insurance sector depend on the insurance market regime. The market regimes were identified by analysing weekly rates of return of the insurers in question during the period between January 2005 and December 2018. They were assessed using statistical clustering methods of time units (weeks) to which there had been assigned conditional variances obtained from the estimated copula-DCC-GARCH model. Indeed, we assumed that the the change (increase) of the risk (variance) is a good (and, moreover, classical) index of the financial market tension. The advantage of such an approach is that there is no need to assume a priori a number of market regimes, because the latter is identified by the clustering quality assessment. Next, in each of the identified regimes we

established the CoVaR systemic risk measure, commonly used nowadays (see e.g. [Acharya et al. 2010], [Bierth et al. 2015], [Jobst 2014]). We assumed that the European insurance market is represented by the weekly rates of return from the STOXX 600 Europe Insurance index. The CoVaR measure, indicating the contribution of each of the insurers to systemic risk, was assessed using the conditional distributions obtained from eight bivariate copula-DCC-GARCH models. In each of these models one boundary distribution represents the European insurance market (the logarithmic return from the STOXX 600 Europe Insurance index), while the other one represents the insurer (the appropriate logarithmic rate of return). To the best of our knowledge, such an approach has not been used in systemic risk analysis ever before.

The paper is organised as follows. The second chapter reviews the subject literature devoted to identification of of systemic risk in the insurance sector, the third chapter presents the methodology and the empirical strategy used in the paper, the fourth one presents the data and discusses the results obtained, while the fifth chapter contains the conclusions.

**2. Systemic risk in insurance – a literature review**

Let us start with recalling a natural definition of systemic risk which is "any set of circumstances that threatens the stability of or public confidence in the financial system" [Billio et al. 2012].

Usually, it is endogenous, coming from the financial system itself, and amplifies the exogenous risk. It can be seen as a coordination failure. Specific sources of systemic crisis are contagion, bank runs, liquidity crisis. Up to now, insurance has virtually been immune to systemic risk, which is partly explained by pyramidal risk sharing (that removes a lot of contagion risk) and less room for coordination failure than in other financial institutions. However, as insurance companies become increasingly involved in other financial activities, or rather, as insurance is more and more often carried out by financial institutions that do not specialize only in this sector, the situation may well change. Of course, there are other causes that may lead to this, such as e.g. more pervasive liquidity insurance offer by the companies. In particular, these conclusions can be found in the Special Report [Geneva Association 2010], Systemic risk in insurance – An analysis of insurance and financial stability. Also in the work [Billio et al. 2010] already mentioned the growing interrelations between the insurance, banking, hedge funds sectors are cited as one of the causes of increasing systemic risk.

Another question is how to measure systemic risk, as several approaches are possible. Leaving for the moment this question (raised in many of the papers listed below, e.g. [Bernardi, Catania, 2015], let us have a quick look on recent points of view on systemic risk in insurance.

The general and most widespread view is that the contribution of the insurance sector to systemic risk (whatever its definition and measurement tools) is very low (for various reasons), but in recent times this is subject to change as the insurance market keeps evolving (compare also the 2015 *Report on systemic risks in the EU insurance sector* [ESRB, 2015]).

Indeed, M. Kanno in [Kanno, 2016] observes that contrary to the interbank market, the insurance one does not contain the feedback mechanisms that would make it fully interconnected. However, he points out that interconnectedness in the insurance sector has not been explored yet using network theory or contagious default approach. As a conclusion, the author sustains the opinion of [IAIS, 2011]. Insurance and financial stability that the degree of interconnectedness within the (re)insurance sector is small which adds to the immunity to systemic risk. However, an earlier study [Dungey, Luciani, Veredas, 2014] notes that insurance companies display substantial systemic risk via interconnectednees with the financial sector and the real economy. Similarly, in [Bierth, Irresberger, Weiß, 2015] the authors studied the contribution of 253 international life and non-life insurers to global systemic risk in the period from 2000 to 2012, observing that systemic risk in the international insurance sector is small in comparison to the case of banks. Still, during the financial crisis insurers did contribute significantly to the instability of the financial sector. In conclusion, the various factors determining systemic risk of insurers are an insurer's interconnectedness and leverage, loss ratios, and the insurer's funding fragility. They also state that there is no big difference in the contribution to global systemic risk between life insurers and non-life insurers. In particular, there seems to be no relations between an insurer's size and it contribution. The authors support the point of view that unlike in the banking sector, the insurance one predominantly suffers from being exposed to systemic risk, rather than adding to the financial system's fragility. Also the study [Mühlnickel, Weiß, 2015] indicates a strong positive relation between consolidation in the insurance industry and moderate systemic risk in the insurance sector, but definitely no extreme systemic risk. Similar conclusion are drawn by Elia Berdin, Matteo Sottocornola (using three measurements) in [Berdin, Sottocornola, 2015]: the insurance industry has a persistent systemic relevance over time, but far from the role of banks in causing systemic risk compared to banks. An interesting contrast between the Eurozone and the USA is observed by Oscar Bernal, Jean-Yves Gnabo, Grégory Guilmin in [Bernal, Gnabo, Guilmin, 2014] where the authors conclude that in the Eurozone during the years 2004-2012, the other financial services sector and the banking sector contribute relatively more to systemic risk in periods of distress than the insurance sector, while in the USA the insurance industry is the systemically riskiest financial sector.

These recent results were preceded by several articles (many of them triggered off by the AIG collapse in the recent crisis) in the years 2009-2013 (as listed in the excellent survey [Eling, Pankoke, 2014]). Scott E. Harrington in [Harrington, 2009] claimed that traditional insurance products do not contribute to systemic risk. Marc P. Radice in [Radice, 2010] came to a two-fold conclusion, namely that unavailability of insurance, insurance run on life insurers, CDS payment default, credit rating utilization (long-term investment, short-term funded) do not contribute to systemic risk, whereas asset contagion, limited fungibility of available group liquidity, distress of non-regulated/non-insurance business within an insurance group could be systemically risky. Faisal Baluch, Stanley Mutenga, Chris Parsons in [Baluch, Mutenga, Parsons, 2011] observed that systemic risk in insurance has grown in the last years, since insurers increased their participation in the capital markets and offered more banking services. The same year Iman van Lelyveld, Franka Liedorp, Manuel Kampman in [van Lelyveld, Liedorp, Kampman, 2011] studied contagion and the contribution of the linkages between insurers and reinsurers to systemic failure coming to the conclusion that even if many reinsurers would go bankrupt the market would not fail and only a few primary insurers would go bankrupt – the potential failure of a (re)insurer (or several of them) is thus not a systemic risk. Still in 2011, a study of the US insurance sector was performed by J. David Cummins, Mary A. Weiss [Cummins, Weiss, 2014a] showing that traditional insurers' activities do not contribute to systemic risk, it is the derivatives trading and financial guarantees that might contribute: life insurers are vulnerable due to leverage, both life and property-casualty insurers are vulnerable to reinsurance crisis; this was corroborated in 2013 again by the same authors in [Cummins, Weiss, 2014b] – they claimed that core property-casualty insurance and reinsurance activities do not contribute to systemic risk, as opposed to non-core insurance activities such as trading in derivatives, asset lending and management, financial guarantees. Martin F. Grace stated in [Grace, 2011] that insurers do not contribute to systemic risk, since duration of assets and liabilities are more closely matched than in the case of banks. Similarly, Denis Kessler in [Kessler, 2013] asserted that reinsurance does not contribute to systemic risk; Patrizia Baur, Rudolf Enz, Aurelia Zanetti in [Baur, Enz, Zanetti, 2003] had come to the same conclusion. On the other hand, in [Mühlnickel,Weiß, 2014] the authors claim that insurers can contribute to systemic risk and are vulnerable to impairments of the financial system. D. Schwarcz and S. L. Schwarcz in [Schwarcz, Schwarcz, 2014] concentrate on systemic risk in insurance as resulting from correlations among firms.

Let us end with some general remarks. Andreas A. Jobs in [Jobst, 2014] also underlines that in general core insurance activities are unlikely to cause or propagate systemic risk. Nevertheless, he suggests that further study is needed with a particular focus on non-traditional and/or non-insurance activities. The interdependencies in the international financial market are a major determinant of systemic risk; they are analysed in e.g. [Brechmann et al., 2013], [Reboredo, Ugolini, 2015], [Di Bernardino et al., 2015]. Note that the problem of assessing the risk, of finding good risk measures, still remains a major research problem, see e.g. [Barrieu et al., 2014] and [Tang, Yang, 2012]. Among the most recent copula-based approaches let us cite [Di Clemente, 2018], [Karimalis, Nomikos, 2018] and [Oh, Patton, 2018].

## 3. Methodology

The empirical strategy we are using in this article in order to analyze the dependances and assess systemic risk on the European insurance market consists of two basic steps:
1. Market regime identification,
2. Analysis, in the identified market regimes, of:
   - the dependances between the studied insurance companies,
   - the correlations between a given insurance company and the European insurance market as represented by the STOXX 600 Europe Insurance index,
   - the systemic risk.

It is assumed in the first step that market regimes are identified using statistical methods of grouping weekly periods $t$ according to the assigned conditional variances of the rates of return of all the instruments being analyzed. The conditional variances that are essential in this approach are obtained through the multivariate copula-DDC-GARCH model. In this model the distribution of the rates of return vector $r_t = (r_{1,t},...,r_{k,t})'$, conditional with respect to the set $\Omega_{t-1}$ of informations available up to the moment $t-1$, is modelled using the conditional copulae proposed by Patton [2006]. It takes the following form:

$$r_{1,t} | \Omega_{t-1} \sim F_{1,t}(\cdot | \Omega_{t-1}),...,r_{k,t} | \Omega_{t-1} \sim F_{k,t}(\cdot | \Omega_{t-1}) \tag{1}$$

$$r_t | \Omega_{t-1} \sim F_t(\cdot | \Omega_{t-1}) \tag{2}$$

$$F_t(r_t | \Omega_{t-1}) = C_t\big(F_{1,t}(r_{1,t} | \Omega_{t-1}),...,F_{k,t}(r_{k,t} | \Omega_{t-1})\big) \tag{3}$$

where $C_t$ denotes the copula, whereas $F_t$ and $F_{i,t}$ are the multivariate CDF and the CDFs of the marginal distributions at time $t$, respectively. In the general case, the univariate rates of return $r_{i,t}$ can be modelled by various specifications of the mean model (e.g. the ARIMA process) and various specifications of the variance model (e.g. sGARCH, fGARCH, eGARCH, gjrGARCH, apARCH, iGARCH, csGARCH).

In our study the following ARIMA process is applied for all series of returns for the mean:

$$r_{i,t} = \mu_{i,t} + y_{i,t}, \tag{4}$$

$$\mu_{i,t} = E(r_{i,t} | \Omega_{t-1}), \quad \mu_{i,t} = \mu_{i0} + \sum_{j=1}^{P_i} \varphi_{ij} r_{i,t-j} + \sum_{j=1}^{Q_i} \theta_{ij} y_{i,t-j} \tag{5}$$

$$y_{i,t} = \sqrt{h_{i,t}} z_{i,t}, \tag{6}$$

while for the variance we use the eGARCH model [Nelson 1991]:

$$\log(h_{i,t}) = \omega_i + \sum_{j=1}^{p_i} \left( \alpha_{ij} \varepsilon_{i,t-j} + \gamma_{ij} \left( |\varepsilon_{i,t-j}| - E|\varepsilon_{i,t-j}| \right) \right) + \sum_{j=1}^{q_i} \beta_{ij} \log(h_{i,t-j}), \quad \varepsilon_{i,t} = \frac{y_{i,t}}{\sqrt{h_{i,t}}}, \tag{7}$$

where $z_{i,t}$ are independent random variables with the same distribution (in the empirical analysis we considered the following distributions: normal, skew-normal, t-Student, skew-t-Student and GED).

The structure of the dependances between the rates of return is modelled using elliptic copulae with conditional correlations $R_t$ as parameters, the dynamics of which is described by the DCC($m$, $n$) model:

$$H_t = D_t R_t D_t, \tag{8}$$

$$D_t = diag(\sqrt{h_{1,t}},...,\sqrt{h_{k,t}}), \tag{9}$$

$$R_t = (diag(Q_t))^{-1/2} Q_t (diag(Q_t))^{-1/2}, \tag{10}$$

$$Q_t = \left(1 - \sum_{j=1}^{m} c_j - \sum_{j=1}^{n} d_j\right) \overline{Q} + \sum_{k=1}^{m} c_j (\varepsilon_{t-j} \varepsilon'_{t-j}) + \sum_{k=1}^{n} d_j Q_{t-j}, \tag{11}$$

where the conditional variances $h_{i,t}$ are modelled using univariate GARCH($p$,$q$) processes of the form (7), $\varepsilon_t = D_t^{-1} y_t$ ($y_t = (y_{1,t},...,y_{k,t})'$) and $\overline{Q}$ is the unconditional covariance matrix of the standardized residuals $\varepsilon_t$. In the specification (11) $c_j$ ($j = 1,...,m$), $d_j$ ($j = 1,...,n$) are scalars describing the influence on the current correlations of the respective previous shocks and previous conditional correlations.

The parameters of the copula-DCC-GARCH model above are estimated using the *inference function for margins - IFM* approached. This method is presented in details e.g. in: [Joe 1997, s. 299–307], [Doman 2011, s. 35–37], [Wanat 2012, s. 98-99]. The computations were done in the R environment using the "rmgarch" package developed by Alexios Ghalanos.

To identify market regimes statistical methods of unsupervised classification are used. We assumed that the groups obtained form periods *t* with similar levels of risk (i.e. having a similar conditional variance). The clustering is performed by means of hierarchical methods in which groups are created recursively by connecting the most similar objects (Ward's method) We are also using two division methods, i.e. the classical k-means method and the partitioning around medoids method (PAM) proposed by Kaufman and Rousseeuw [1990]. The optimal number of groups, and thus the market regimes, are assessed under the following measures of cluster validity: the *Calinski-Harabasz index* [Calinski, Harabasz 1974], the *silhouette index- SI*) [Kaufman, Rousseeuw 1990], the *Dunn index*) [Dunn 1974] and the *Xie-Beni separation measure*) [Xie, Beni 1991].

In the second stage of analysis, in each of the identified market regimes we assessed the *CoVaR*. The systemic risk measure $CoVaR_{\beta,t}^{i|j}$ is defined to be the value at risk (*VaR*) of the institution (market index) *i* under the condition that another institution (market index) *j* is subject to distress i.e. its rate of return is smaller than its value at risk, meaning:

$$P\left(r_{i,t} \leq CoVaR_{\beta,t}^{i|j} \mid r_{j,t} \leq VaR_{\alpha,t}^{j}\right) = \beta \qquad (12)$$

Using the conditional probability formula we get:

$$\frac{P\left(r_{i,t} \leq CoVaR_{\beta,t}^{i|j}, r_{j,t} \leq VaR_{\alpha,t}^{j}\right)}{P\left(r_{j,t} \leq VaR_{\alpha,t}^{j}\right)} = \beta \qquad (13)$$

The definition of the value at risk for the institution *j*, i.e. $VaR_{\alpha,t}^{j}$ yields $P\left(r_{j,t} \leq VaR_{\alpha,t}^{j}\right) = \alpha$, that is:

$$P\left(r_{i,t} \leq CoVaR_{\beta,t}^{i|j}, r_{j,t} \leq VaR_{\alpha,t}^{j}\right) = \alpha\beta. \qquad (14)$$

Therefore, the assessment of $CoVaR_{\beta,t}^{i|j}$ requires knowledge of the bivariate distribution $F_t$ of the vector $(r_{i,t}, r_{j,t})$. Due to the Sklar Theorem this distribution can be represented using the copula in the following way:

$$F_t(r_{i,t}, r_{j,t}) = C_t(F_i(r_{i,t}), F_j(r_{j,t})). \qquad (15)$$

Invoking (15), $CoVaR_{\beta,t}^{i|j}$ can be determined by solving (numerically) the equation:

$$C_t(F_i(CoVaR_{\beta,t}^{i|j}), \alpha) = \alpha\beta. \qquad (16)$$

In the empirical analysis we studied the influence on the European insurance market's systemic risk of the five largest insurance companies from Europe and the biggest insurers from the USA, Canada and China. In accordance, it was assumed that $r_{i,t}$ represents the European insurance market (we made use of the weekly rates of return from STOXX 600 Europe Insurance), while $r_{j,t}$ describes the insurers (we made use of the weekly logarithmic returns on shares). For each of the eight pairs:

(*rate of return from the* STOXX 600 *index* $r_{i,t}$, *logarithmic return of the insurer* $r_{j,t}$)

we assessed the parameters of the bivariate copula-DCC-GARCH model described by the formulae (1)-(7). Then, using these parameters together with the conditional correlations obtained by these models, we determined the copulae $C_t$ and the distributions $F_t$. The values $CoVaR_{\beta,t}^{i|j}$ for the analyzed period were obtained by solving numerically the equation (16).

## 4. Data and analysis results

As a basis for our study we took the stock prices of the five largest insurers from Europe and the biggest insurers from the USA, Canada and China (cf. Tab. 1 and Fig. 1) as well as the STOXX 600 Europe Insurance index representing the European insurance market (cf. Fig. 2). We analyzed the weekly weekly log-returns for the period between January 2005 and December 2018.

Table 1. Insurance companies considered in the study with their acronyms used in the presentation of the results

| No. | Insurer | Acronym | Country | Total assets (in mld USD) |
|---|---|---|---|---|
| 1 | AXA | AXA | France | 944.145 |
| 2 | Allianz | Allianz | Germany | 934.654 |
| 3 | Prudential plc | Prud | Great Britain | 578.149 |
| 4 | Legal & General | Legal | Great Britain | 574.901 |
| 5 | Aviva | Aviva | Great Britain | 541.188 |
| 6 | Metlife | Metlife | USA | 898.764 |
| 7 | Manulife Financial | Manu | Canada | 534.705 |
| 8 | Ping An Insurance | Ping | China | 802.975 |

Source: Compiled from http://www.relbanks.com/top-insurance-companies/world

In the first stage of study we identified the regimes of the insurance market on the basis of the conditional variances of the rates of return of the insurance companies in question. These were assessed using the 8-variate copula-DCC-GARCH model. During the analysis we considered various ARMA-GARCH specifications of univariate models. Finally, on the grounds of information criteria and model appropriateness tests (result available upon request to the author) we opted, for all the instruments, for the ARMA(1,1)-eGARCH(2,2) model with the skew Student distribution (with skewness $\xi$ and shape $\upsilon$)[1]. During the analysis of the dynamics of the dependances between the rates of return we considered the Gauss and Student copulae together with various specifications of the DCC model. As earlier, in the basis of information criteria we chose the Student copula with conditional correlation coefficients obtained from the DCC(1, 1) model and a constant shape parameter $\eta$. The assessment results are presented in Table 2, while the conditional variances obtained are shown in Figure 3.

Figure 1. Quotations of the insurance companies studied for the period 07.01.2005 - 21.12.2018

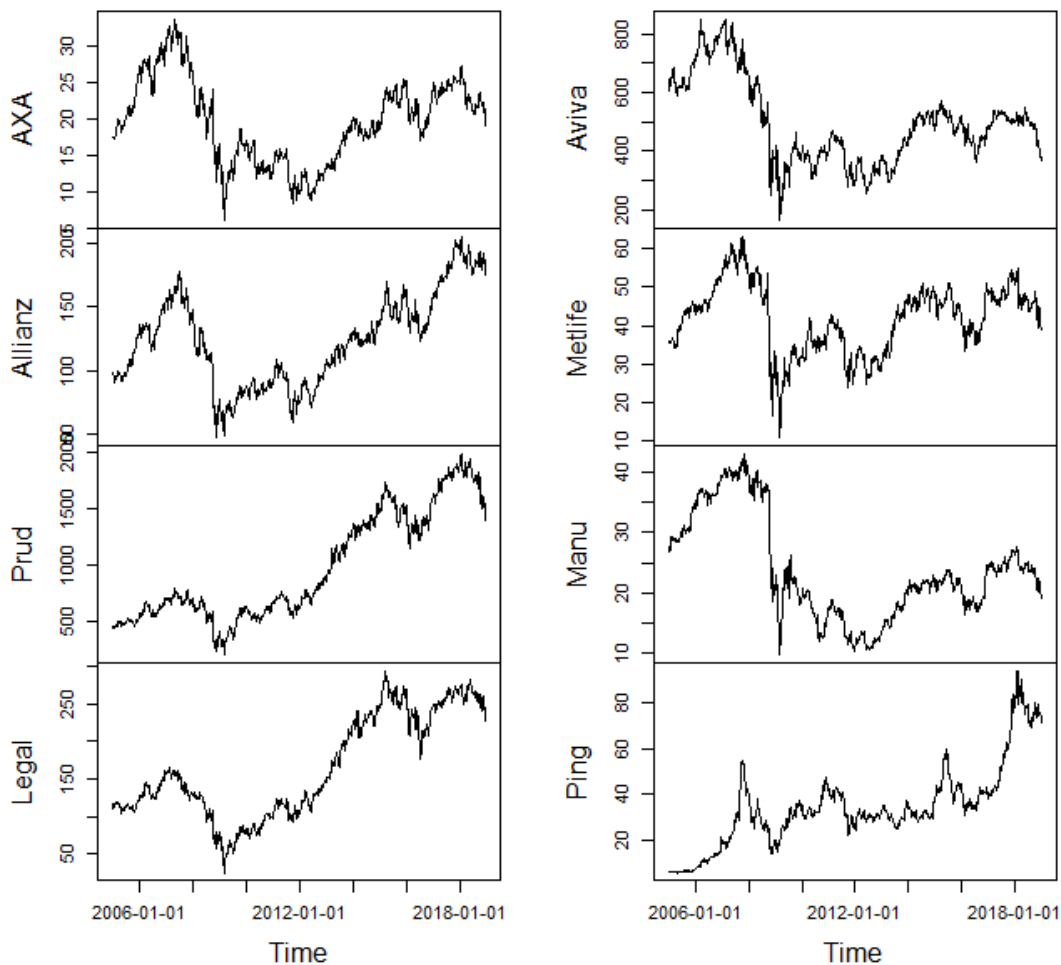

*Source*: Author's own study.

---
[1] The eGARCH means exponential GARCH model put forward by Nelson.

Figure 2. STOXX 600 Europe Insurance index during the period 07.01.2005 - 21.12.2018

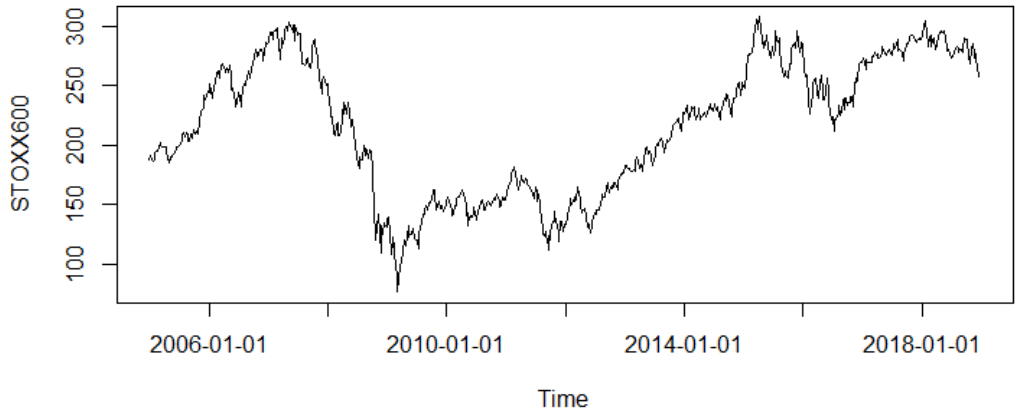

*Source*: Author's own study.

Table 2. Copula-DCC–GARCH model estimation results

| Param. | AXA | Allianz | Prud | Legal | Aviva | Metlife | Manu | Ping |
|---|---|---|---|---|---|---|---|---|
| $\mu$ | 0.00096 | 0.00109 | 0.00088 | 0.00110 | -0.00059 | 0.00090 | 0.00025 | 0.00394 |
|  | *0.35692* | *0.32940* | *0.28148* | *0.06039* | *0.56085* | *0.09304* | *0.80474* | *0.01980* |
| $\varphi_1$ | 0.84452 | 0.28441 | 0.60722 | 0.74250 | 0.72107 | 0.78764 | -0.86865 | -0.93666 |
|  | *0.00000* | *0.00000* | *0.00000* | *0.00000* | *0.00000* | *0.00000* | *0.00054* | *0.00000* |
| $\theta_1$ | -0.88966 | -0.33966 | -0.73356 | -0.81227 | -0.77971 | -0.84399 | 0.80562 | 0.91175 |
|  | *0.00000* | *0.00000* | *0.00000* | *0.00000* | *0.00000* | *0.00000* | *0.00730* | *0.00000* |
| $\Omega$ | -0.18911 | -0.20240 | -0.12869 | -0.18544 | -0.24923 | -0.16082 | -0.20182 | -0.27178 |
|  | *0.00001* | *0.00008* | *0.00106* | *0.02815* | *0.00398* | *0.00409* | *0.01234* | *0.08481* |
| $\alpha_1$ | -0.30001 | -0.25969 | -0.19735 | -0.20584 | -0.17208 | -0.19634 | -0.18676 | -0.03959 |
|  | *0.00000* | *0.00004* | *0.00028* | *0.01911* | *0.00025* | *0.00000* | *0.00971* | *0.36170* |
| $\alpha_2$ | 0.20207 | 0.15644 | 0.09066 | 0.10362 | -0.01844 | 0.08099 | 0.02249 | 0.03560 |
|  | *0.00018* | *0.01251* | *0.10526* | *0.32087* | *0.73485* | *0.00843* | *0.72404* | *0.48038* |
| $\beta_1$ | 1.00000 | 1.00000 | 1.00000 | 1.00000 | 0.19314 | 1.00000 | 0.59507 | 0.24303 |
|  | *0.00000* | *0.00000* | *0.00000* | *0.00000* | *0.00000* | *0.00000* | *0.00000* | *0.00000* |
| $\beta_2$ | -0.02985 | -0.03002 | -0.02046 | -0.02843 | 0.76824 | -0.02518 | 0.37509 | 0.71127 |
|  | *0.00002* | *0.00014* | *0.00090* | *0.03756* | *0.00000* | *0.00443* | *0.00000* | *0.00000* |
| $\gamma_1$ | -0.03655 | 0.05317 | -0.09294 | 0.12954 | 0.02767 | 0.11685 | 0.14136 | 0.28382 |
|  | *0.63396* | *0.60211* | *0.28044* | *0.20640* | *0.72923* | *0.25838* | *0.15328* | *0.00001* |
| $\gamma_2$ | 0.17901 | 0.05073 | 0.21587 | 0.07644 | 0.30621 | 0.01034 | 0.08238 | 0.09320 |
|  | *0.03338* | *0.63029* | *0.01862* | *0.46385* | *0.00006* | *0.91989* | *0.33795* | *0.20801* |
| $\xi$ (skew.) | 0.85192 | 0.83317 | 0.80222 | 0.89056 | 0.81522 | 0.87085 | 0.92185 | 1.13212 |
|  | *0.00000* | *0.00000* | *0.00000* | *0.00000* | *0.00000* | *0.00000* | *0.00000* | *0.00000* |
| $v$ (shape) | 11.73223 | 10.13238 | 6.04081 | 5.43739 | 6.06003 | 4.51676 | 5.06490 | 5.48191 |
|  | *0.01179* | *0.00788* | *0.00003* | *0.00000* | *0.00000* | *0.00000* | *0.00001* | *0.00000* |
| Copula-DCC parameters | | | | | | | | |
| Distribution | Octovariate t-Student | | | | | | | |
| DCC order | DCC(1.1) | | | | | | | |
|  | Parameters | | | | | | | |
| $c_1$ | 0.01063 (*0.00012*) | | | | | | | |
| $d_1$ | 0.94801 (*0.00000*) | | | | | | | |
| $\eta$ (shape) | 9.96436 (*0.00000*) | | | | | | | |

Probability values (p-values) are in parentheses.

*Source: Author's own calculation.*

Market regimes were identified by means of clustering weekly periods with respect to the conditional variances of the rates of return of the studied insurance companies. In this crucial – from the point of view of the whole study – step we considered various combinations of distance measures, clustering methods and number of classes. Eventually, led by criteria of clustering quality (cf. Table 3) we chose a division into two classes obtained using the method of k-means with the Euclidean distance (cf. Fig. 4). In this case the silhouette index is 0.8683 (clustering quality is pictured in Fig. 5). We assumed therafter that to different classes there correspond different market regimes. The variance distribution in different regimes is shown in Fig. 6. From it we can infer that the first regime is characterized by a low volatility (low risk level), while the second one, occurring during the period 17.10.2008-05.06.2009 – a high volatilty (high risk level).

Figure 3. Conditional variances

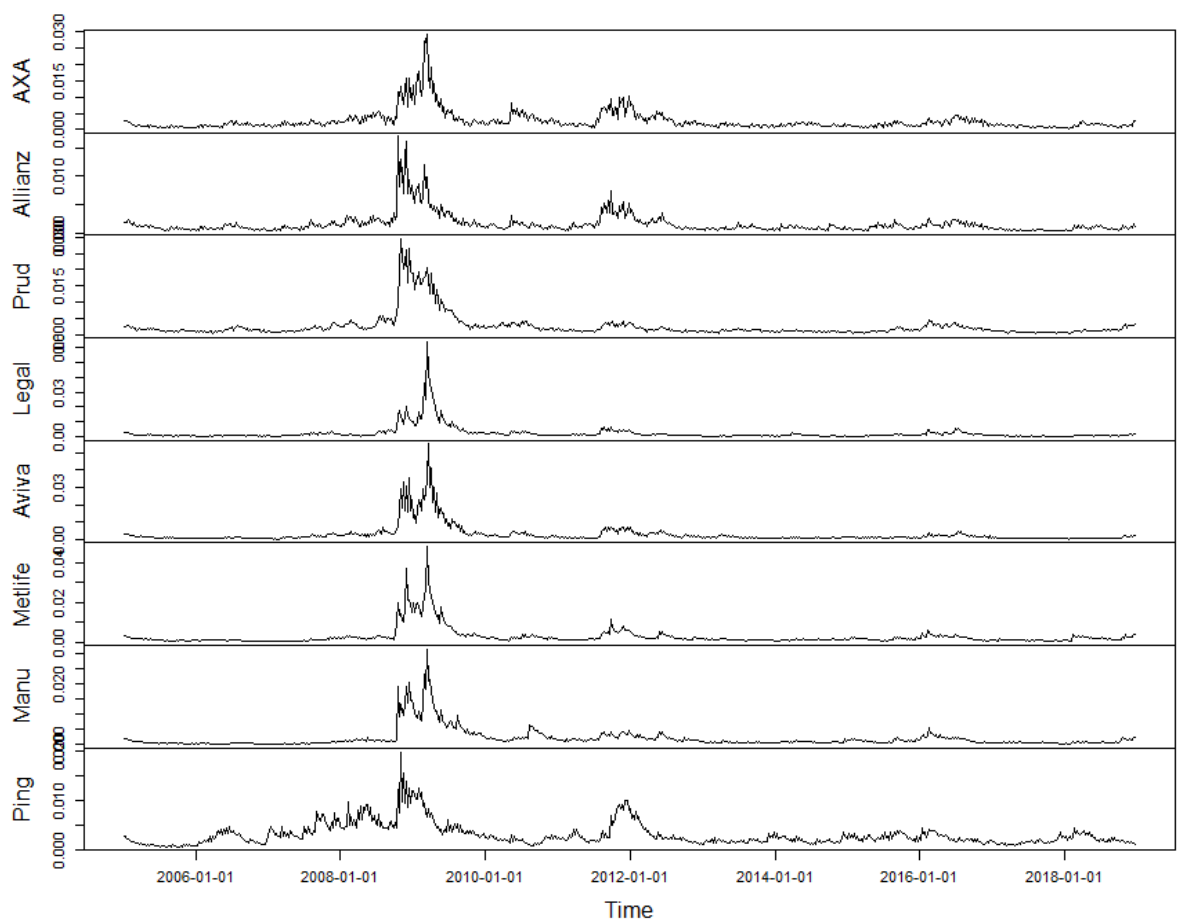

*Source: Author's own calculation.*

Table 3. Validation indices for data partitions

| Validation criterion | Number of clusters | | | | |
|---|---|---|---|---|---|
| | 2 | 3 | 4 | 5 | 6 |
| | Ward's method | | | | |
| Silhouette | **0.8683** | 0.4202 | 0.3958 | 0.3987 | 0.3986 |
| Calinski Harabasz index | **1545.1570** | 1006.8530 | 771.5901 | 963.3596 | 814.7552 |
| Dunn index | **0.0552** | 0.0080 | 0.0080 | 0.0110 | 0.0110 |
| Xie-Beni index | **1.9208** | 76.1650 | 68.5520 | 45.4610 | 43.3223 |
| | PAM | | | | |
| Silhouette | **0.8623** | 0.4788 | 0.4153 | 0.4181 | 0.1549 |
| Calinski Harabasz index | **1501.2950** | 1036.3830 | 791.2769 | 990.6590 | 809.8822 |
| Dunn index | **0.0353** | 0.0082 | 0.0077 | 0.0104 | 0.0053 |
| Xie-Beni index | **4.1444** | 66.2503 | 72.2987 | 47.7384 | 177.4645 |
| | k-means | | | | |
| Silhouette | **0.8683** | 0.5238 | 0.5177 | 0.4713 | 0.4394 |
| Calinski Harabasz index | **1545.1570** | 1063.6570 | 1170.1440 | 1047.2740 | 915.3568 |
| Dunn index | **0.0552** | 0.0071 | 0.0106 | 0.0146 | 0.0127 |
| Xie-Beni index | **1.9208** | 92.8171 | 62.4426 | 28.7042 | 34.8416 |

*Source*: *Author's own calculation.*

Note: numbers in bold indicate the optimal number of groups with reference to a given criterion.

Figure 4. Identified market regimes

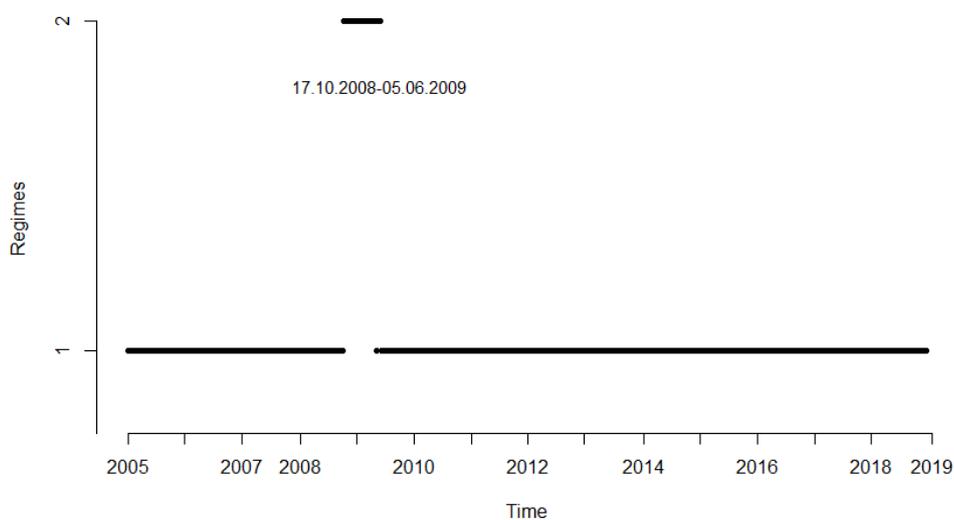

*Source*: *Author's own calculation*

Figure 5. Silhouette plot

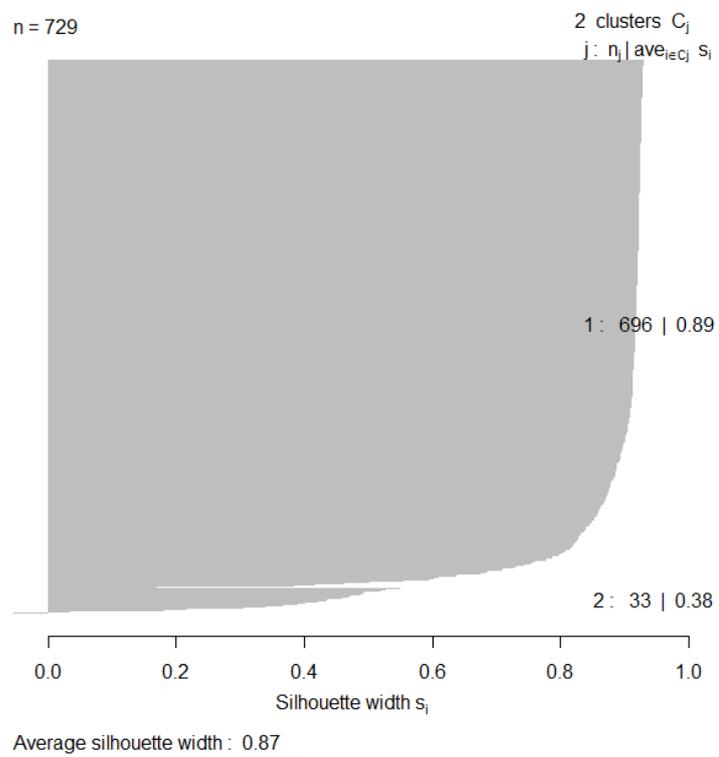

*Source*: *Author's own calculation.*

Figure 6. Distribution of the conditional variance in the identified market regimes

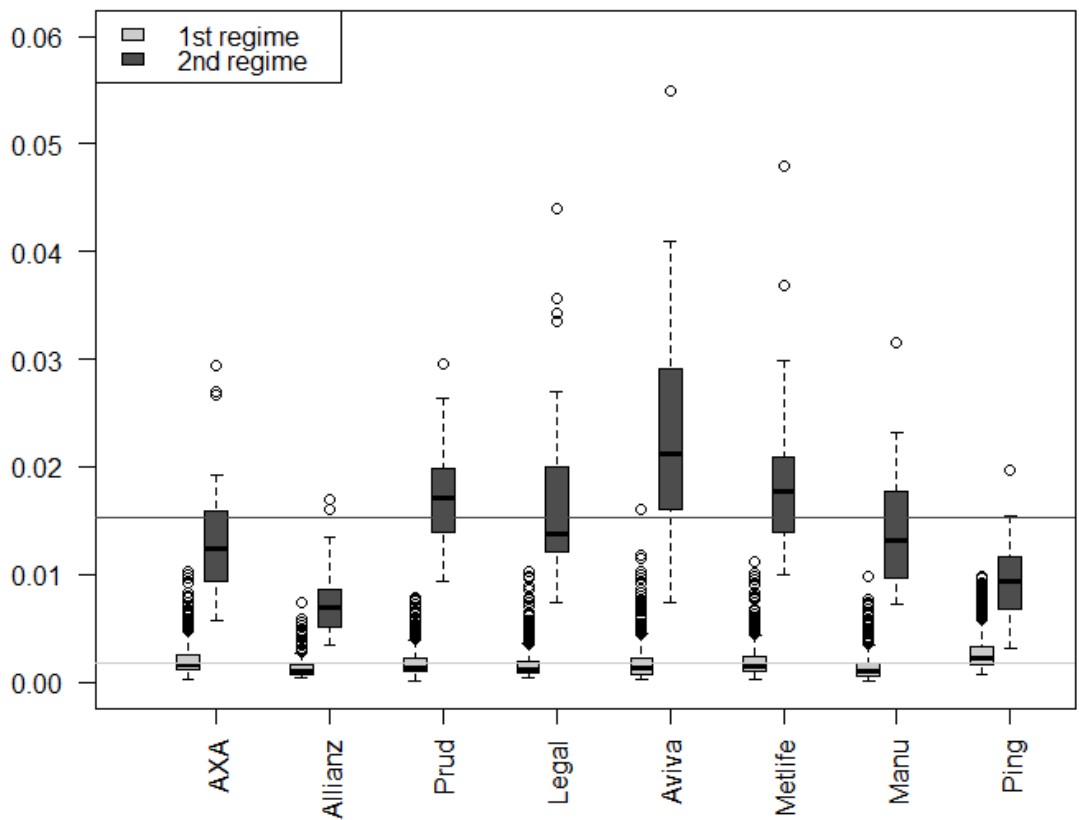

*Source*: *Author's own calculation.*

In the second step of our studies the analysis of the dependances between the studied insurance companies was done based on the conditional correlations from the previously assessed octovariate copula-DCC-GARCH model. Their distribution for the respective pairs in the identified market regimes is shown in Figure 7.

On the other hand, the analysis of the dependances between the insurer and the European insurance market, as well as the analysis of systemic risk in the first and second market regime was carried out on the basis of the estimated eight bivariate copula-DCC-GARCH models for the following pairs: the rate of return on the European market index and the individual rate of return for the given insurance company. In the case of the insurers we were working with the earlier estimated ARMA(1,1)-eGARCH(2,2) models with the skew Student distribution. On the grounds of information criteria and model appropriateness tests we considered the same specification for the STOXX 600 Europe Insurance index rate of return. The parameters of the estimated model are given in Table 4. During the analysis of the dynamics between the rate of return on the index representing the European insurance market and the insurers' rates of return, we were considering the Gauss and Student copulae as well as various specifications of the DCC model. On the basis of information criteria for each pair we chose the Student copula with conditional correlations obtained from the DCC(1, 1) model and constant shape parameters. The estimation results are presented in Table 5, while the conditional correlations obtained are shown in Figure 8. Finally, the distribution of the conditional correlations between the domestic and European capital markets in the identifie regimes is given in Figure 9.

The systemic risk assessment in the identified market regimes was performer using the *CoVaR* measure determined by the method described in the previous section. The *CoVaR* value distribution illustrating the influence of a given insurer on the European insurance market is shown in Figure 10.

Figure 7. Distribution of the conditional correlations between analyzed markets in the identified regimes

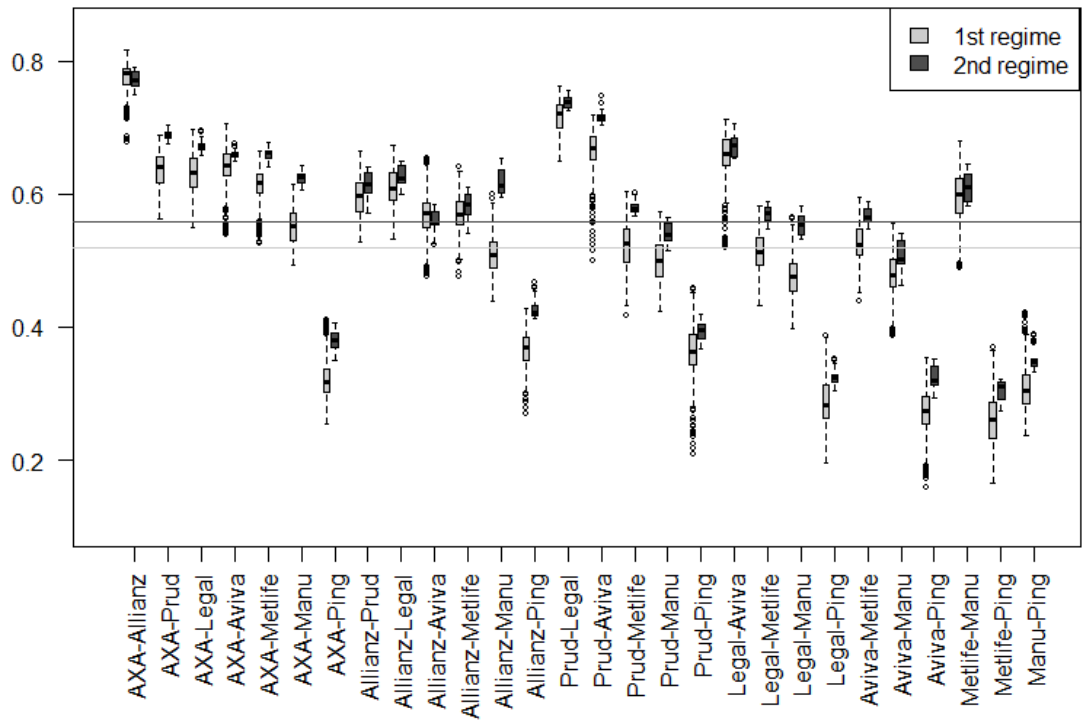

*Source*: Author's own study

Table 4. Univariate ARAM(1,1)- eGARCH(2,2) model estimations for the STOXX 600 Europe Insurance index

| Parameter | $\mu$ | $\varphi_1$ | $\theta_1$ | $\omega$ | $\alpha_1$ | $\alpha_2$ |
|---|---|---|---|---|---|---|
| estimation | 0.00086 | 0.68392 | -0.72741 | -0.20234 | -0.25814 | 0.16812 |
| p-Value | *0.35844* | *0.00000* | *0.00000* | *0.00221* | *0.00003* | *0.00570* |
| Parameter | $\beta_1$ | $\beta_2$ | $\gamma_1$ | $\gamma_2$ | $\xi$ (skew.) | $\upsilon$ (shape) |
| estimation | 1.00000 | -0.02848 | 0.09708 | 0.05358 | 0.79261 | 9.87665 |
| p-Value | *0.00000* | *0.00691* | *0.35870* | *0.60707* | *0.00000* | *0.00438* |

*Source*: Author's own calculations

Table 5. Bivariate DCC(1, 1) models estimations for the pairs: STOXX 600 Europe Insurance and a given insurer

|  | AXA | Allianz | Prud | Legal | Aviva | Metlife | Manu | Ping |
|---|---|---|---|---|---|---|---|---|
| $c_1$ | 0.02513 | 0.02159 | 0.03199 | 0.04218 | 0.02631 | 0.07105 | 0.03338 | 0.00942 |
|  | *0.04014* | *0.01083* | *0.02421* | *0.00990* | *0.00191* | *0.02998* | *0.08039* | *0.73776* |
| $d_1$ | 0.95214 | 0.96262 | 0.94015 | 0.92320 | 0.96805 | 0.72663 | 0.90777 | 0.85545 |
|  | *0.00000* | *0.00000* | *0.00000* | *0.00000* | *0.00000* | *0.00027* | *0.00000* | *0.32845* |
| $\eta$ (shape) | 6.85867 | 11.11860 | 8.13343 | 7.99758 | 6.31898 | 16.80825 | 7.86310 | 15.97758 |
|  | *0.00026* | *0.00241* | *0.00000* | *0.00012* | *0.00000* | *0.10873* | *0.00122* | *0.14136* |

*Source:* Author's own calculations

Figure 8. Conditional correlations between the insurer and the European insurance market

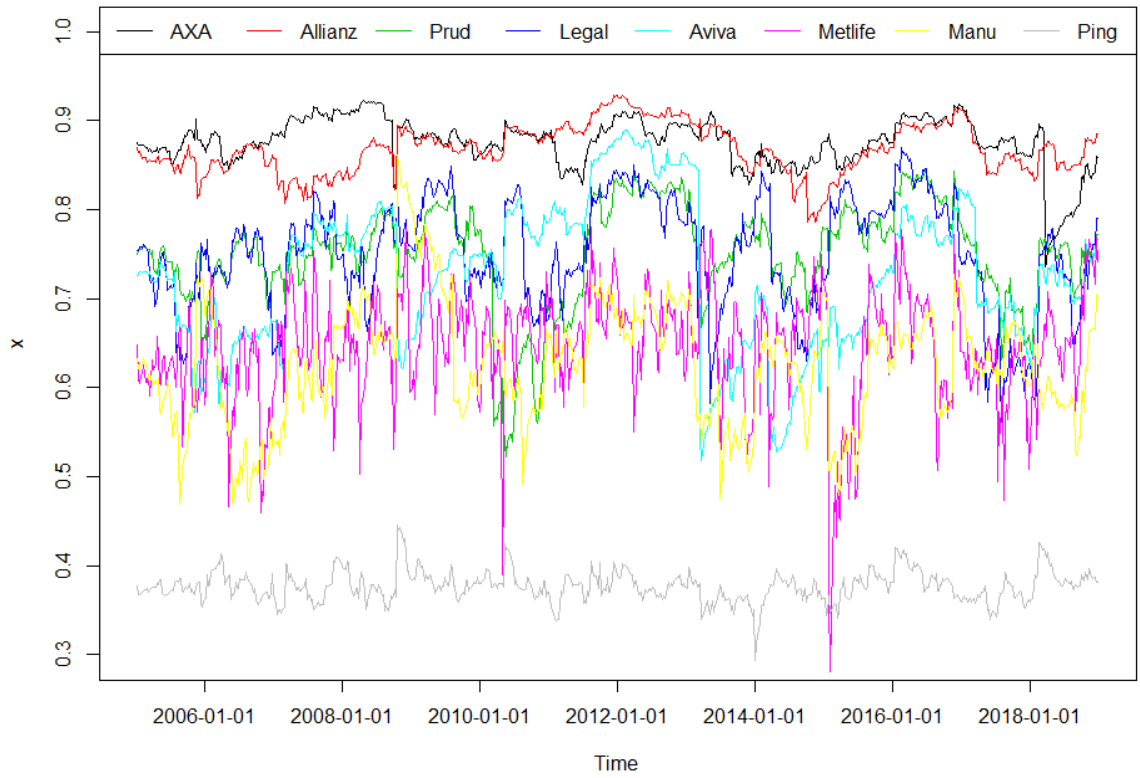

*Source*: *Author's own calculation.*

Figure 9. Distribution of the conditional correlations between the insurer and the European insurance market in the identified regimes

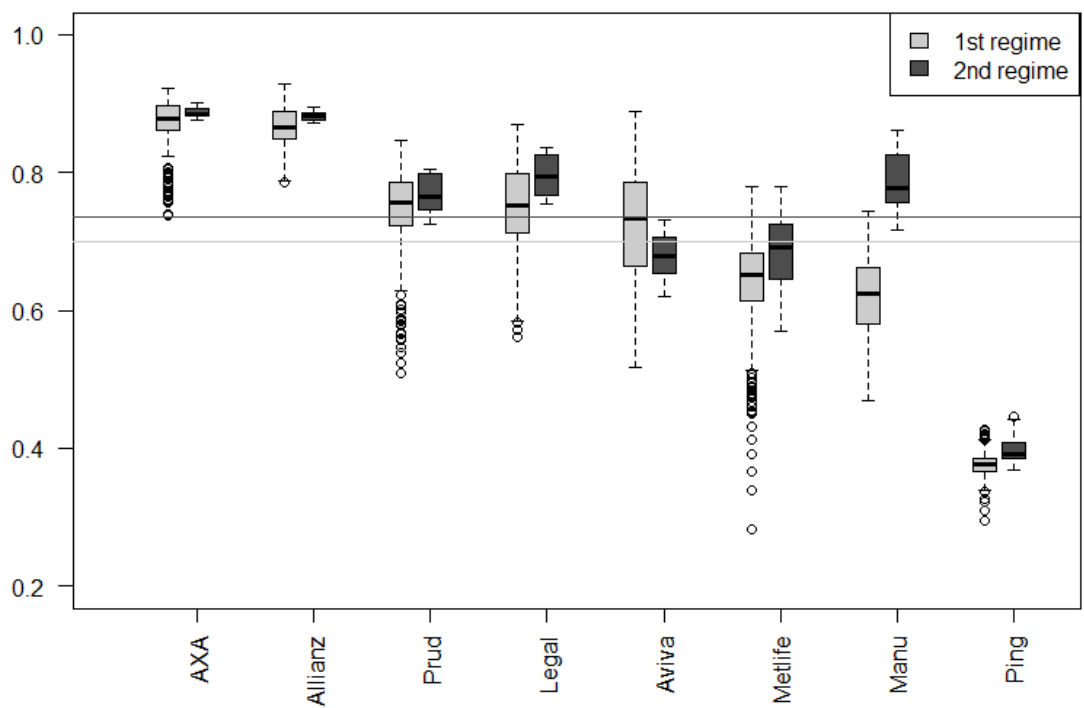

*Źródło*: Opracowanie własne

Figure 10. Distribution of the *CoVaR* measure in the identified regimes

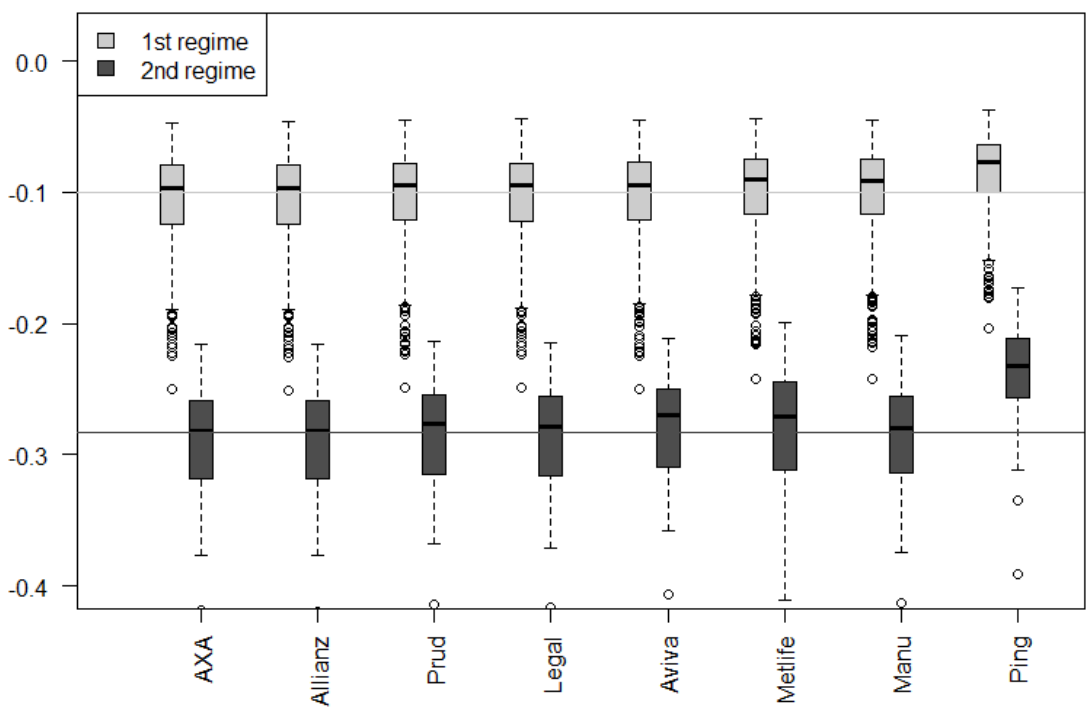

*Source*: Author's own work

## 5. Conclusions

In this work we used the copula-DCC-GARCH model to analyze the dependances in the group formed by the largest five insurance companies from Europe and the biggest insurers from the USA, Canada and China. Then, availing ourselves of the *CoVaR* measure we studied the influence of each insurer on the European insurance market systemic risk. The European market was represented by the STOXX 600 Europe Insurance index, while for the insurers we considered their quotations on domestic markets. The study was performer in two steps. The first one consisted in identifying the regimes of the European insurance market, while in the second one, we analyzed – for the identified regimes – the following items: the correlations (usisng conditional correlations) between the considered insurance companies, the dependances between a given insurer and the European insurance market as well as the influence of the studied insurance companies on the European insurance market systemic risk. The market regimes were identified by monitoring the insurers' logarithmic returns on shares. To this aim we applied statistical clustering methods for weekly periods to which we were assigning the conditional variances obtained from the estimated octovariate copula-DCC-GARCH model. Both, the clustering quality measures and the possibility of a reasonable economical intepretation exposed two different market regimes in the considered period of time: a regime of low volatility (1st regime – `normal') and a regime of unstable quotations (2nd regime – `risk') that appeared during the time of the biggest turbulences experienced by the global markets.

We can draw the following conclusions from our study:

– The insurance companies from the investigated group are positively correlated. The strongest dependance is to be seen among the insurers from Europe (Axa and Allianz are the pair with the strongest tie), a somewhat weaker dependance exists between the insurers from Europe and those from North America, while the weakest is between the insurer from China and the remaining ones. These correlations are clearly stronger in the second identified regime, i.e. during the turbulences on global markets period (cf. Fig. 7). On that basis we can state that during a global crisis the exposure to systemic risk on the European insurance market increases.

– The European insurance market as represented by the STOXX 600 Europe Insurance index is most strongly correlated to the largest insurance companies from Europe, i.e. Axa or Allianz, a weaker correlation exists in the case of insurers from North America and a notably weaker still in the case of the insurer from China[2] (cf. Fig. 8). As earlier, these correlations are sronger in the second market regime (cf. Fig. 9).

- There is an important difference between the *CoVaR* measures for the first and second regimes of the European insurance market in the case of all the insurers from the studied group. The influence of insurance companies on systemic risk is much stronger during the turbulences period (cf. Fig. 10). It is also apparent that in a fixed regime this influence is more or less at the same level, which in the case of the insurer from China is somewhat lower than average.
- The influence of insurance companies from North America on the European inurance market systemic risk is at a comparative level with the influence of companies from Europe, both in the first and second identified market regimes.

---

[2] It should be noted here that these results may be biased to some extend by the construction of the STOXX 600 Europe Insurance index

Cracow University of Economics

Department of Mathematics

Rakowicka 27

31-510 Kraków

Poland

wanats@uek.krakow.pl, anna.denkowska@uek.krakow.pl